\documentclass[pra,twocolumn,floatfix,a4paper,superscriptaddress]{revtex4}

\usepackage{bm,color,amsmath,txfonts}
\usepackage{graphicx}
\usepackage{siunitx}
\usepackage{subfigure}
\usepackage{verbatim}
\usepackage{dcolumn}
\usepackage{bm}
\usepackage{epsf}
\usepackage{xcolor}
\usepackage{hyperref}
\usepackage{hhline}
\usepackage{float}
\usepackage{enumerate}
\usepackage{bbm}
\usepackage{lipsum}
\usepackage{mathrsfs}
\usepackage{ulem}

\begin{document}

\title{Entangling two exciton modes using exciton optomechanics}

\author{Xuan Zuo}
\affiliation{Interdisciplinary Center of Quantum Information, State Key Laboratory of Modern Optical Instrumentation, and Zhejiang Province Key Laboratory of Quantum Technology and Device, School of Physics, Zhejiang University, Hangzhou 310027, China}
\author{Zhi-Yuan Fan}
\affiliation{Interdisciplinary Center of Quantum Information, State Key Laboratory of Modern Optical Instrumentation, and Zhejiang Province Key Laboratory of Quantum Technology and Device, School of Physics, Zhejiang University, Hangzhou 310027, China}
\author{Huai-Bing Zhu}
\affiliation{Interdisciplinary Center of Quantum Information, State Key Laboratory of Modern Optical Instrumentation, and Zhejiang Province Key Laboratory of Quantum Technology and Device, School of Physics, Zhejiang University, Hangzhou 310027, China}
\author{Jie Li}\thanks{jieli007@zju.edu.cn}
\affiliation{Interdisciplinary Center of Quantum Information, State Key Laboratory of Modern Optical Instrumentation, and Zhejiang Province Key Laboratory of Quantum Technology and Device, School of Physics, Zhejiang University, Hangzhou 310027, China}

\begin{abstract}
Exciton optomechanics, bridging cavity exciton polaritons and optomechanics, opens new opportunities for the study of light-matter strong interactions and nonlinearities, due to the rich nonlinear couplings among excitons, phonons, and photons. Here, we propose to entangle two exciton modes in an exciton-optomechanics system, which consists of a semiconductor microcavity integrated with two quantum wells. {The quantum wells support two exciton modes, which simultaneously couple to an optical cavity mode via a linear dipole interaction and to a mechanical vibration mode via a nonlinear deformation potential interaction.} We show that by strongly driving the microcavity with a red-detuned laser field and when the two exciton modes are respectively resonant with the Stokes and anti-Stokes sidebands scattered by the mechanical motion, stationary entanglement between the two exciton modes can be established under realistic parameters. {The protocol is within reach of current technology and may become a promising approach for preparing excitonic entanglement.}
\end{abstract}

\maketitle

\section{Introduction}

Semiconductor optomechanical microcavities integrated with quantum wells (QWs) can simultaneously confine photons, phonons and excitons in a compact mode volume and achieve strong coupling among them~\cite{review,EOM1, EOM2, EOM3}. This novel hybrid system is termed as exciton optomechanics~\cite{review}.
The strong coupling between excitons and microcavity photons leads to exciton polaritons~\cite{Hopfield58,Weisbuch92}, a type of half-matter half-light bosons with a very small effective mass~\cite{Keeling07, Deng10}. In the past decades, significant progress has been made in exciton polaritons in both fundamental physics and practical applications, e.g., Bose-Einstein condensation~\cite{BEC1, BEC2}, light emitting diodes~\cite{Tsintzos08}, and low-threshold room-temperature polariton lasers~\cite{Imamoglu96, Christopoulos07, Kena-Cohen10} have been achieved.     

The dual light-matter nature of exciton polaritons also leads to a tunable and strongly enhanced polariton-mechanics coupling~\cite{Barg18, Kobecki22}, because the exciton-phonon interaction can be much stronger~\cite{Bir,EOM3} than the two types of the optomechanical interaction, i.e., radiation pressure~\cite{MA14} and photoelasticity~\cite{Perrin13, Favero14}. A near-unity single-polariton quantum cooperativity is promising to be achieved using current semiconductor resonator platforms~\cite{EOM3}. Moreover, in such a semiconductor configuration, exciton polaritons can be efficiently pumped by the electrical current, which offers an additional effective means to control the system dynamics~\cite{Schneider13, Bhattacharya13, Bhattacharya14}, and the exciton component can bring in rich nonlinearities to the system, including the exciton-exciton and -phonon interactions. Meanwhile, highly monochromatic phonons at gigahertz can be generated by bulk acoustic wave resonators and effectively injected into the microcavity to modulate the polaritonic states~\cite{Santos21}. Coherent mechanical self-oscillation can be achieved via the polariton drive and the phonon laser~\cite{Fainstein20} and the microwave-optical conversion~\cite{Santos23} have been realized in this hybrid system. 

Exciton optomechanics exhibits a wealth of unexplored physics and prospective applications that can be uncovered by combining quantum optics and semiconductor physics. However, so far research in this field has predominantly focused on the classical phenomena, and the quantum effects have been rarely explored~\cite{review,EOM3,Zuo24}. Here, we present a quantum theory for achieving stationary entanglement of two exciton modes by exploiting the dispersive {exciton-phonon interaction via the deformation potential} in an exciton-optomechanics system, which contains two QWs supporting two exciton modes. 
{Specifically, when the microcavity is driven by a strong red-detuned laser and when the two exciton modes are respectively resonant with the two sidebands scattered by the mechanical motion, both the Stokes and anti-Stokes scattering are resonantly enhanced. The former entangles excitons with phonons through a parametric down conversion (PDC) interaction, and the latter cools the phonon mode and meanwhile transfers the generated exciton-phonon entanglement to the two exciton modes via the phonon-exciton state-swap interaction. Therefore, the two exciton modes are entangled due to the mediation of the phonon mode in the above two simultaneous scattering processes.} 
The entanglement is in the steady state and robust against various dissipations of the system and bath temperature.

The paper is organized as follows. In Sec.~\ref{model}, we introduce the exciton-optomechanics system and provide its Hamiltonian and the corresponding quantum Langevin equations (QLEs). We further show how to linearize the dynamics and achieve the steady-state entanglement of the two exciton modes. In Sec.~\ref{result}, we analyze the mechanism of entanglement generation, present the results of the exciton-exciton entanglement, and discuss the impact of the couplings and dissipations of the system on the entanglement.  Finally, we conclude in Sec.~\ref{conc}.

\begin{figure}[b]
	\includegraphics[width=0.96\linewidth]{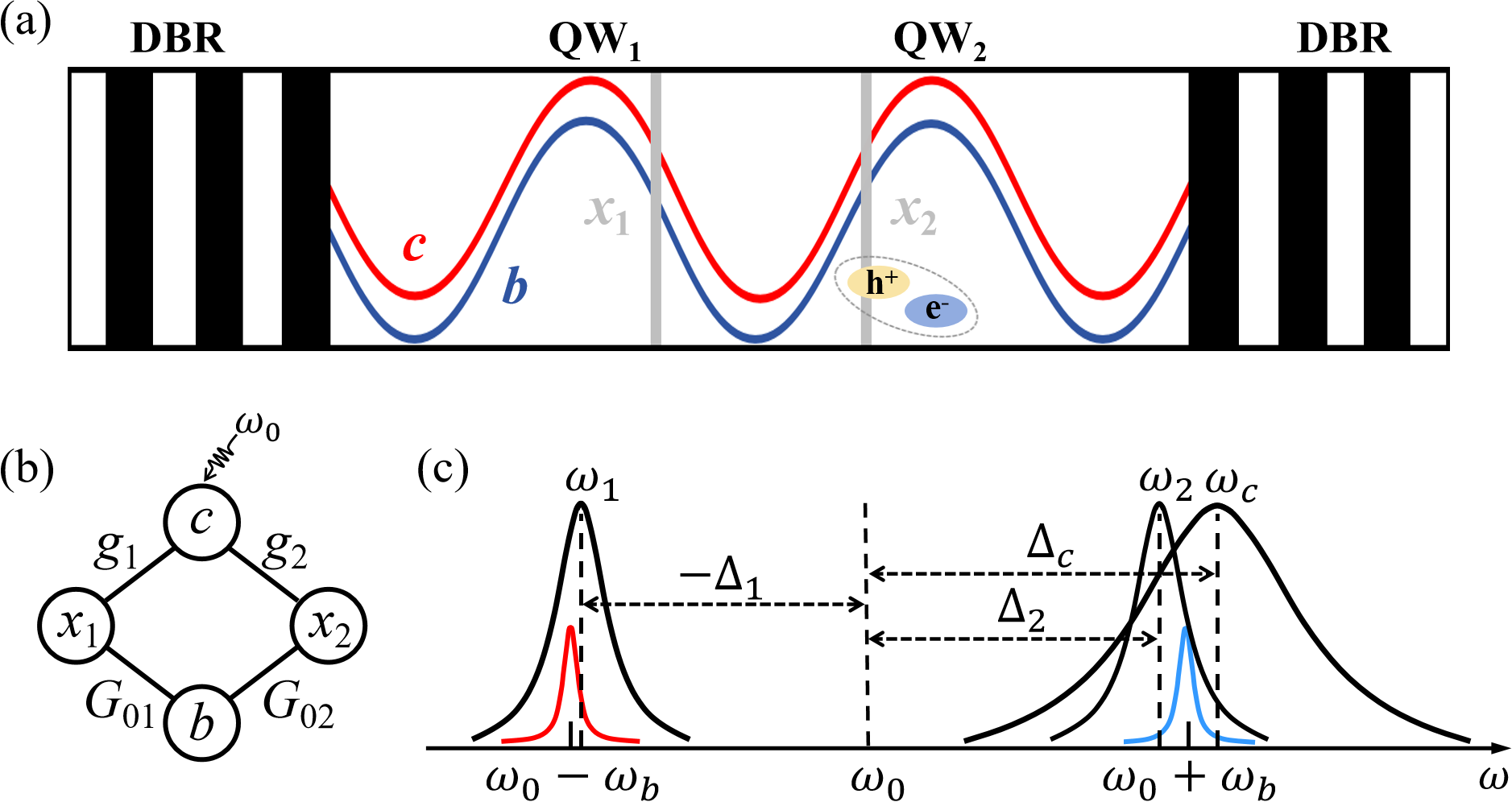}
	\caption{(a) Sketch of the exciton-optomechanics system. {A semiconductor microcavity formed by two movable DBRs confines a cavity mode $c$ and a mechanical mode $b$. Two QWs, supporting two exciton modes $x_1$ and $x_2$, are placed away from (close to) the antinodes of the cavity field (the strain field). (b) The cavity mode $c$, driven by a laser with frequency $\omega_0$, couples to the two exciton modes $x_1$ and $x_2$ via a linear beam-splitter interaction, and the two exciton modes further couple to the mechanical mode $b$ via a dispersive electrostrictive interaction. (c) Frequencies and linewidths of the system. Two sidebands at $\omega_0\pm\omega_b$ are produced due to the scattering of the mechanical motion.} When the cavity mode with frequency $\omega_c$ is resonant with the anti-Stokes sideband and the two exciton modes with frequencies $\omega_{1,2}$ are respectively resonant with the two mechanical sidebands, the two exciton modes get entangled.}
	\label{fig1}
\end{figure}

\section{The model}\label{model}

We consider a hybrid exciton-optomechanics system~\cite{EOM1,EOM2,EOM3}, which consists of a semiconductor microcavity sandwiched between two distributed Bragg reflectors (DBRs) and two QWs that are placed {away from} the antinodes of the cavity field, as depicted in Fig.~\ref{fig1}(a). The DBRs are constructed from layers with alternating high and low refractive indices and act as high-reflectivity mirrors to effectively confine optical photons. Each QW supports an exciton mode, where electrons and holes form dipoles capable of interacting with an electromagnetic field, embodied by a linear beam-splitter-type exciton-photon interaction. Due to the radiation-pressure and the photoelastic interaction, the microcavity layer and DBRs vibrate, promising a dispersive optomechanical coupling of the cavity mode with the vibration phonon mode~\cite{Perrin13,Favero14}. 
{In addition, semiconductor materials offer an alternative mechanism for coupling with the mechanical mode through the exciton-mediated electrostrictive force, i.e., the exciton mode can couple to the mechanical mode via the deformation potential interaction, which is also a dispersive interaction as the strain in the material perturbs the semiconductor band structure, yielding a frequency shift of the exciton mode~\cite{Bir}. Both the theory~\cite{EOM3} and experiment~\cite{Sesin} indicate that the exciton-phonon coupling strength can be two orders of magnitude larger than the photon-phonon coupling strength in the semiconductor microcavities. So, here we primarily consider the strong exciton-phonon dispersive interaction, while neglect the much weaker optomechanical interaction. Note that this corresponds to the configuration where the QWs are placed close to the antinodes of the strain field~\cite{review,Sesin}, which is the derivative of the displacement distribution. The QWs are also placed away from the antinodes of the cavity field for a relatively weak exciton-photon coupling to avoid forming exciton polaritons because we aim to entangle two exciton modes.}
The Hamiltonian of such a quadripartite system, cf. Fig.~\ref{fig1}(b), reads
{
\begin{align}\label{HHH}
\begin{split}
H/\hbar &= \sum_{k=1,2} \omega_{k} x_k^\dagger x_k + \omega_{c} c^\dagger c + \omega_b b^\dagger b + i\Omega \left(c^\dagger e^{-i\omega_0t}-{\rm H.c.} \right)
\\& + \sum_{k=1,2} g_k \left(x_k^\dagger c + x_k c^\dagger \right) + G_{\rm 0k} x_k^\dagger x_k \left(b +  b^\dagger \right),
\end{split}
\end{align}}
where $x_1$, $x_2$, $c$ and $b$ ($x_1^\dagger$, $x_2^\dagger$, $c^\dagger$ and $b^\dagger$) are the annihilation (creation) operators of the two exciton modes, the cavity mode, and the mechanical mode, respectively, satisfying $[j,j^{\dagger}]=1$ ($j=x_1,\,x_2,\,c,\,b$); $\omega_i$ ($i=1,2,c,b$) are their resonance frequencies; $g_k$ denotes the coupling strength between the cavity with the $k$th exciton mode, which can be much stronger than the cavity decay rate $\kappa_c$ and the exciton dissipation rate $\kappa_k$, resulting in the formation of exciton polaritons~\cite{Hopfield58,Weisbuch92}; {$G_{\rm 0k}$ represents the bare exciton-phonon coupling strength}; the last term in the first line is the driving Hamiltonian, where $\Omega = \sqrt{2 P \kappa_c/ \hbar \omega_0}$ denotes the coupling strength between the drive field and the microcavity, with $\omega_0$ ($P$) being the frequency (power) of the driving laser.

Including the dissipations and input noises of the system, we obtain the following QLEs in the frame rotating at the drive frequency $\omega_0$:
{
\begin{align}\label{QLExcb}
	\begin{split}
		\dot{x}_1=&-(i\Delta_1 + \kappa_1) x_1 - i g_1 c - i G_{01} x_1 ( b + b^\dagger ) + \sqrt[]{2\kappa_1}x_1^{in},  \\
		\dot{x}_2=&-(i\Delta_2 + \kappa_2) x_2 - i g_2 c - i G_{02} x_2 ( b + b^\dagger ) + \sqrt[]{2\kappa_2}x_2^{in},  \\
		\dot{c}=&-(i\Delta_c + \kappa_c) c - i \!\sum_{k=1,2} g_k x_k  + \Omega + \sqrt[]{2\kappa_c}c^{in}, \\
		\dot{b}=&-(i\omega_b + \kappa_b) b - i \!\sum_{k=1,2} G_{\rm 0k} x_{k}^\dagger x_{k} + \sqrt[]{2\kappa_b} b^{in} ,
	\end{split}
\end{align}}
where $\Delta_k=\omega_k-\omega_0$ ($k=1,2$) and $\Delta_c=\omega_c-\omega_0$; $\kappa_b$ is the mechanical damping rate, and $j^{in}(t)$ are the input noise operators for the mode $j$ ($j=x_1,\,x_2,\,c,\,b$), which are zero mean and characterized by the correlation functions~\cite{Zoller}: $\langle j^{in}(t)j^{in\dagger}(t^\prime) \rangle=[N_j(\omega_j)+1]\delta(t-t^\prime)$, $\langle j^{in\dagger}(t)j^{in}(t^\prime) \rangle=N_j(\omega_j)\delta(t-t^\prime)$, with $N_j(\omega_j)=[\exp[(\hbar \omega_j/k_BT)]-1]^{-1}$ being the equilibrium mean thermal excitation number of the mode $j$, and $k_B$ as the Boltzmann constant and $T$ the bath temperature.

The cavity is strongly driven by an intense laser field {and due to the photon-exciton excitation-exchange interaction, the two exciton modes have a large amplitude $| \langle x_k \rangle| \gg 1$ at the steady state.} This allows us to linearize the system dynamics around the large average values by writing each mode operator $j$ as its average $\langle j \rangle$ plus the quantum fluctuation operator $\delta j$, i.e., $j = \langle j \rangle \,+\, \delta j$, and neglecting small second-order fluctuation terms. Consequently, we obtain two sets of equations for the classical averages and the quantum fluctuations, respectively. The former set of equations are given by {$\langle x_1 \rangle = -i g_1 \langle c \rangle/(i \tilde{\Delta}_1 + \kappa_1)$, $\langle x_2 \rangle = -i g_2 \langle c \rangle/(i \tilde{\Delta}_2 + \kappa_2)$, $\langle b \rangle \approx -\frac{1}{\omega_b} \left(G_{01} | \langle x_1 \rangle |^2 + G_{02} | \langle x_2 \rangle |^2 \right)$, and
\begin{align}\label{stSol}
	\begin{split}
		\langle c \rangle  = \frac{\Omega (i \tilde{\Delta}_1 + \kappa_1)(i \tilde{\Delta}_2 + \kappa_2)}{g_1^2 (i \tilde{\Delta}_2 {+} \kappa_2) \,{+} g_2^2 (i \tilde{\Delta}_1 {+} \kappa_1) \,{+}\, (i {\Delta}_c {+} \kappa_c) (i \tilde{\Delta}_1 {+} \kappa_1) (i \tilde{\Delta}_2 {+} \kappa_2)},
	\end{split}
\end{align}
where $\tilde{\Delta}_1 = {\Delta}_1 + 2 G_{01} \langle b \rangle$ and $\tilde{\Delta}_2 = {\Delta}_2 + 2 G_{02} \langle b \rangle$ are the effective exciton-drive detuning, including the frequency shift caused by the deformation potential.} The linearized QLEs for the quantum fluctuations are
{
\begin{align}\label{QLEs}
	\begin{split}
		\dot{\delta x}_1 = & - \big( i \tilde{\Delta}_1 + \kappa_1 \big) \delta x_1 - i g_1 {\delta c} - G_{1} \big( \delta b + \delta b^\dagger \big) + \sqrt[]{2\kappa_1} x_1^{in},  \\
		\dot{\delta x}_2 = & - \big( i \tilde{\Delta}_2 + \kappa_2 \big) \delta x_2 - i g_2 {\delta c} - G_{2} \big( \delta b + \delta b^\dagger \big) + \sqrt[]{2\kappa_2} x_2^{in},  \\
		\dot{\delta c} = & - \big( i {\Delta}_c + \kappa_c \big) \delta c  - i \!\sum_{k=1,2} g_k {\delta x}_k + \sqrt[]{2\kappa_c} c^{in},  \\
		\dot{\delta b} = & - \big( i {\omega}_b + \kappa_b \big) \delta b -  \!\sum_{k=1,2} \big(G_{k} \delta x_k^\dagger - G_{k}^* \delta x_k \big) + \sqrt[]{2\kappa_b} b^{in},
	\end{split}
\end{align}
with $G_1 = i G_{01} \langle x_1 \rangle$ and $G_2 = i G_{02} \langle x_2 \rangle$ being the effective exciton-phonon coupling strength}, which are generally complex.

The above QLEs~\eqref{QLEs} can be expressed in a concise matrix form in terms of the quadrature fluctuations $\delta X_j = (\delta j + \delta j^{\dagger})/\!\sqrt{2}$, $\delta Y_j = i (\delta j^{\dagger} - \delta j)/\!\sqrt{2}$ and the quadrature form of the input noises $X_{j}^{in}$ and $Y_{j}^{in}$ defined analogously, i.e.,
\begin{align}
\dot{u}(t)=  A \, u(t) + n(t),
\end{align}
where $u(t)=[\delta X_{x_1}(t),\delta Y_{x_1}(t),\delta X_{x_2}(t),\delta Y_{x_2}(t),\delta X_c(t),\delta Y_c(t)$, $\delta X_b(t),\delta Y_b(t)]^{\rm T}$, $n(t) = [\sqrt[]{2\kappa_1}X_{x_1}^{in},\sqrt[]{2\kappa_1}Y_{x_1}^{in},\sqrt[]{2\kappa_2}X_{x_2}^{in},\sqrt[]{2\kappa_2}Y_{x_2}^{in}$, $\sqrt[]{2\kappa_c}X_c^{in},\sqrt[]{2\kappa_c}Y_c^{in}, \sqrt[]{2\kappa_b}X_b^{in}, \sqrt[]{2\kappa_b}Y_b^{in}]^{\rm T}$, and the drift matrix $A$ is given by
\small{{
\begin{align}
	 A\,{=}\begin{pmatrix}
		-\kappa_1 & \tilde{\Delta}_1 & 0 & 0 & 0 & g_1 & - 2 {\rm Re}G_1 & 0 \\
		-\tilde{\Delta}_1 & -\kappa_1 & 0 & 0 & -g_1 & 0 & - 2 {\rm Im}G_1 & 0\\
		 0 & 0 & -\kappa_2 & \tilde{\Delta}_2 & 0 & g_2 & - 2 {\rm Re}G_2 & 0\\
		 0 & 0 & -\tilde{\Delta}_2 & -\kappa_2 & -g_2 & 0 & - 2 {\rm Im}G_2 & 0\\
		0 & g_1 & 0 & g_2 & -\kappa_c & {\Delta}_c & 0 & 0\\
		-g_1 & 0 & -g_2 & 0 & -{\Delta}_c & -\kappa_c & 0 & 0\\
		0 & 0 & 0 & 0 & 0 & 0 & -\kappa_b & \omega_b \\
		- 2 {\rm Im}G_1 & 2 {\rm Re}G_1 & - 2 {\rm Im}G_2 & 2 {\rm Re}G_2 & 0 & 0 & -\omega_b & -\kappa_b\\
	\end{pmatrix}.
\end{align}}}
Because the quantum noises are Gaussian and the system dynamics is linearized, the steady state of the quantum fluctuations is a continuous variable four-mode Gaussian state, which can be completely characterized by an $8 \times 8$ covariance matrix (CM) $\cal V$, with the entries ${\cal V}_{ij}=\frac{1}{2} \langle u_i(t)u_j(t^\prime) + u_j(t^\prime)u_i(t) \rangle$ $(i,j=1,2,...,8)$, as the first moments of the quadrature fluctuations are zero. The steady-state CM can be straightforwardly achieved by solving the Lyapunov equation~\cite{Vitali07, Parks}.
\begin{align}
	\begin{split}
	A {\cal V} + {\cal V} A^{\rm T} = -D,
	\end{split}
\end{align}
where $D=\mathrm{Diag} \big[\kappa_1(2N_{x_1} {+} 1),\kappa_1(2N_{x_1} {+} 1),\kappa_2(2N_{x_2} {+} 1), \kappa_2(2N_{x_2}$ $+1),\kappa_c(2N_c {+} 1),\kappa_c(2N_c {+} 1),\kappa_b(2N_b  {+} 1),\kappa_b(2N_b {+} 1)\big]$ is the diffusion matrix, which is defined by $\langle n_i(t)n_j(t')+n_j(t')n_i(t) \rangle/2 = D_{ij}\,\delta(t - t')$. With the CM in hand, the entanglement of two exciton modes can then be quantified by calculating the logarithmic negativity~\cite{Ent}, which for Gaussian states is defined as
\begin{align}
	\begin{split}
	E_N \equiv \max[0,-\ln 2 \tilde{\nu}_-],
	\end{split}
\end{align}
where $\tilde{\nu}_- = \min {\rm eig}|i \Omega_2 \tilde{{\cal V}}_{\rm ex}|$ (the symplectic matrix $\Omega_2 = \oplus_{j=1}^2 i \sigma_y$ and $\sigma_y$ is the $y$-Pauli matrix) is the minimum symplectic eigenvalue of the CM $\tilde{{\cal V}}_{\rm ex} = {\cal P} {\cal V}_{\rm ex} {\cal P}$, with ${\cal V}_{\rm ex}$ being the $4 \times 4$ CM of the two exciton modes obtained by removing in $\cal V$ the rows and columns associated with the cavity and mechanical modes, and ${\cal P} = \mathrm{Diag} [1, -1, 1, 1]$ being the matrix that performs partial transposition on the CM.

\section{Exciton-exciton entanglement} \label{result}

 \begin{figure}[b]
	\includegraphics[width=\linewidth]{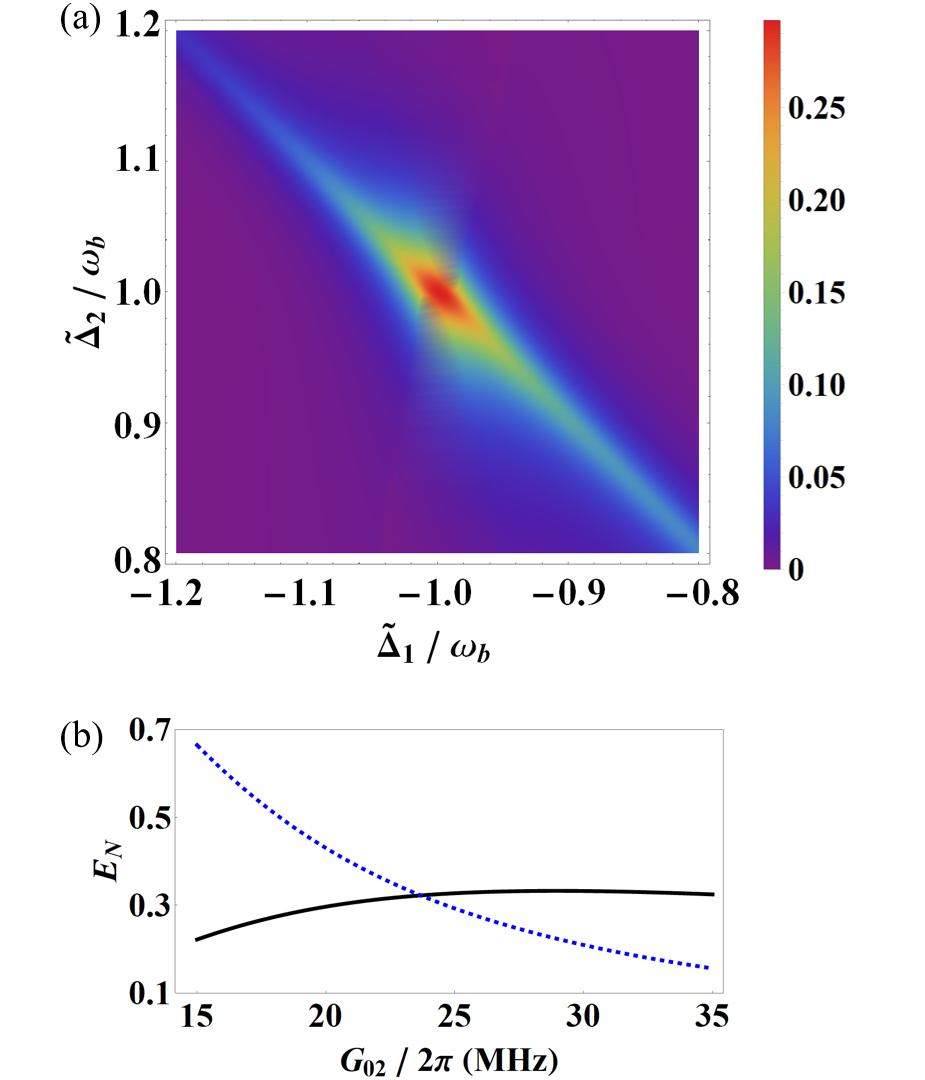}
	\caption{(a) Density plot of the stationary entanglement between two exciton modes versus {two effective exciton-drive detunings $\tilde{\Delta}_1$ and $\tilde{\Delta}_2$.} (b) Stationary exciton-exciton entanglement $E_{x_1,x_2}$ (solid) and {exciton-phonon entanglement $E_{x_1,b}$ (dashed) versus the bare exciton-phonon coupling $G_{02}$. We take ${\Delta}_c = \omega_b$ in (a) and (b), and $\tilde{\Delta}_1 = - \tilde{\Delta}_2 = - \omega_b$ in (b).} The other parameters are provided in the text.}
	\label{fig2}
\end{figure}

In typical semiconductor optomechanical microcavities~\cite{EOM1,EOM3,Perrin13}, the mechanical frequency is high and in the gigahertz range. Although the mean thermal occupation of the mechanical mode is small at low bath temperatures, it becomes large at high temperatures, which inhibits the generation of quantum states. To eliminate this dominant thermal noise of the system, we use a red-detuned laser to drive the microcavity with the detuning $\Delta_c \approx \omega_b$ and {set one exciton mode (e.g., the $x_2$ mode) to be resonant with the cavity, i.e., $\tilde{\Delta}_2 \approx \Delta_c \approx \omega_b$, which resonantly enhances the exciton-phonon anti-Stokes scattering (cf. Fig.~\ref{fig1}(c)). This corresponds to the cooling of the mechanical motion, analogue to the opto-~\cite{MA14} and magnomechanical~\cite{CMMrev} sideband cooling, which optimally works  in the resolved sideband limit $\kappa_2 \ll \omega_b$. }

{
We find that when the other exciton mode (the $x_1$ mode) is further resonant with the Stokes sideband, i.e., $\tilde{\Delta}_1 \approx - \omega_b$ (Fig.~\ref{fig2}(a)), the two exciton modes get entangled. This can be understood in the following way~\cite{CMMrev}.  In the Stokes process, the exciton mode $x_1$ and the phonon mode are entangled due to the PDC interaction; while in the anti-Stokes process, the exciton mode $x_2$ and the phonon mode realize an effective state-swap interaction. Therefore, due to the mediation of the mechanics in the above two processes, the two exciton modes become entangled. This is confirmed by the fact that the exciton-exciton entanglement is maximized at the optimal detunings $\tilde{\Delta}_1 \approx  -\tilde{\Delta}_2 \approx - \omega_b$, as shown in Fig.~\ref{fig2}(a). In Fig.~\ref{fig2}(b), we show the stationary exciton-exciton entanglement $E_{x_1,x_2}$ and the exciton-phonon entanglement $E_{x_1,b}$ versus the bare exciton-phonon coupling strength $G_{02}$. As analyzed above, the entanglement $E_{x_1,b}$ is produced in the Stokes process, whereas the entanglement $E_{x_1,x_2}$ is the result of the transfer of entanglement from the $x_1{\textendash}b$ system to the $x_1{\textendash}x_2$ system through the phonon-exciton $b{\textendash}x_2$ state-swap interaction (with strength $G_{2}$) in the anti-Stokes scattering.
The increasing (decreasing) of the entanglement $E_{x_1,x_2}$ ($E_{x_1,b}$) with the increase of $G_{02}$, i.e., with the increasing effective coupling $G_{2}$ for a fixed drive power, clearly manifests such a process of entanglement transfer. Such an entanglement distribution among different subsystems is a typical feature of the entanglement in multipartite systems~\cite{Li18,LPR}.}
 In getting Fig.~\ref{fig2}, we use experimentally feasible parameters~\cite{review,EOM1,EOM3,Perrin13,Sesin}: $\omega_b/2\pi = 20$ GHz, $\kappa_b/2\pi = 1$~MHz, $\kappa_c/2\pi = {10}$ GHz, $\kappa_{1(2)}/2\pi = 10^2$ MHz, $g_{1(2)}/2\pi = {1}$ GHz, {$G_{01}/2\pi = 10$ MHz}, {$G_{02}/2\pi = 20$ MHz}, $\Omega/2\pi =  {5.5}$ THz, and $T = 1$ K.  Under the optimal detunings in Fig.~\ref{fig2}(a), the mechanical motion is cooled into its quantum ground state with an effective mean phonon number of {0.44}. This corresponds to the drive power $P \approx {1.89}$ mW for $\omega_0 / 2\pi \approx 3 \times 10^2$ THz. {We remark that the frequency of the exciton mode can be adjusted by changing the ratio of the alloy doping so that optimal detunings can be achieved in a practical system.}

\begin{figure}[t]
	\includegraphics[width=\linewidth]{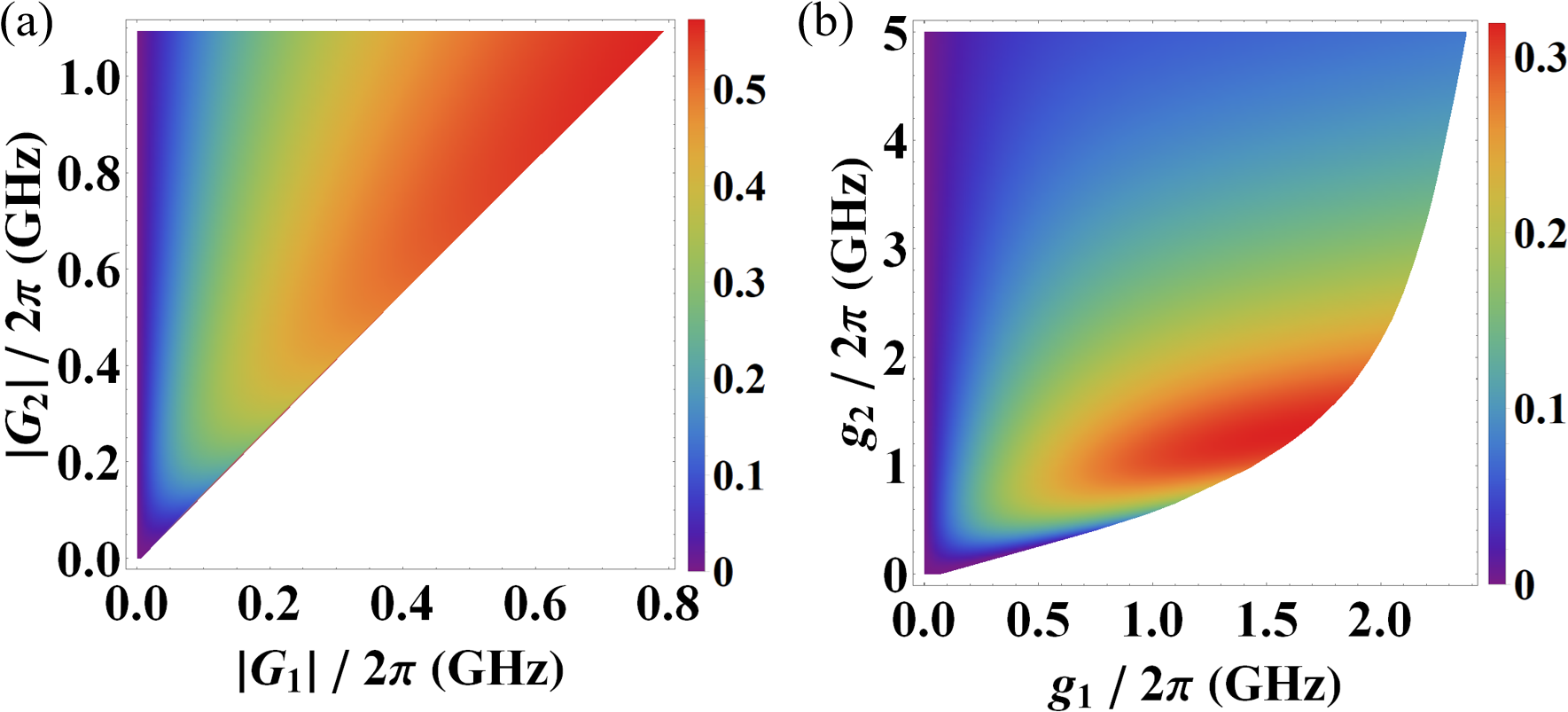}
	\caption{Stationary exciton-exciton entanglement versus (a) {exciton-phonon couplings $G_{1}$ and $G_{2}$};  (b) exciton-photon couplings $g_1$ and $g_2$. We take the optimal detunings {$\Delta_c = \omega_b$ and $\tilde{\Delta}_1 = -\tilde{\Delta}_2 = -\omega_b$. The blank area indicates the regime where the system is unstable.} The other parameters are the same as in Fig.~\ref{fig2}.}
	\label{fig3}
\end{figure}

{
Figure~\ref{fig3}(a) shows the exciton entanglement versus the effective exciton-phonon couplings $|G_{1}|$ and $|G_{2}|$. Clearly, it manifests the indispensable role of the dispersive exciton-phonon interaction in creating the entanglement: the entanglement is present only for nonzero $G_{1}$ and $G_{2}$. Typically, $|G_{2}| >|G_{1}|$ is required for keeping the system being stable and a relatively larger $G_2$ can efficiently transfer the entanglement created in the Stokes process to the two exciton modes.
Figure~\ref{fig3}(b) plots the entanglement versus the two exciton-photon couplings $g_1$ and $g_2$. Since the cavity is driven, the exciton-photon coupling strength $g_k$ determines the pump efficiency associated with the $x_k$ exciton mode (cf. Eq.~\eqref{stSol} and equations above), and thus the effective exciton-phonon coupling $G_k$, because $|G_1/G_2| = |G_{01} \langle x_1 \rangle / G_{02} \langle x_2 \rangle| = g_1/2g_2$, under the parameters of Fig.~\ref{fig2}.  The stability condition $|G_{2}| >|G_{1}|$ implies that $g_1<2g_2$. There is an optimal relative strength between the Stokes scattering, which produces entanglement, and the anti-Stokes scattering, which transfers entanglement, i.e., an optimal ratio of $|G_1/G_2|$, for the exciton-exciton entanglement. This means an optimal ratio of $g_1/g_2$: too small $g_1$ is inefficient to create entanglement, while too large $g_1$ leads the system to be unstable. }

{
We remark that in typical semiconductor QW microcavities, the exciton-photon coupling strength $g$ can be very strong and exceed the exciton and cavity decay rates, thus entering the strong-coupling regime and forming exciton polaritons. However, in this scheme, we intentionally position the QWs away from the antinodes of the cavity field to reduce the exciton-photon coupling strength. Moreover, we adopt a practical large cavity decay rate $\kappa_c/2\pi = 10$ GHz, which keeps the cavity-exciton system well within the weak-coupling regime. The reason to avoid forming exciton polaritons is that the polariton mode decays much faster than the exciton mode, because of the large cavity decay rate. The polariton entanglement will decay rapidly due to the mixing of short-lived cavity photons.}

\begin{figure}[t]
	\includegraphics[width=\linewidth]{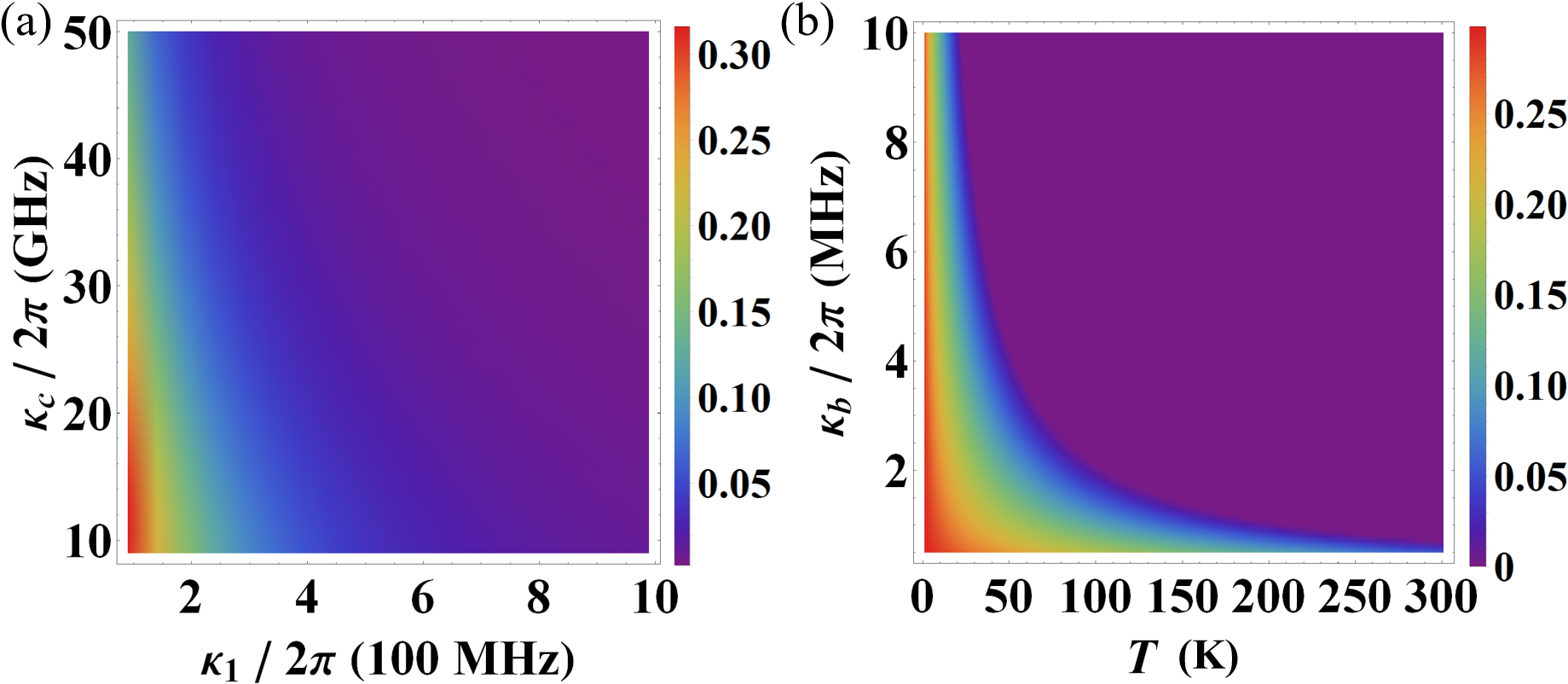}
	\caption{Stationary exciton-exciton entanglement versus (a) exciton and cavity dissipation rates $\kappa_1$ $(= \kappa_2)$ and $\kappa_c$; (b) bath temperature $T$ and mechanical damping rate $\kappa_b$. {We take ${\Delta}_c = \omega_b$ and $\tilde{\Delta}_1 = -\tilde{\Delta}_2 = -\omega_b$.} The other parameters are those of Fig.~\ref{fig2}.}
	\label{fig4}
\end{figure}

In Fig.~\ref{fig4}, we study the impact of various dissipation rates and bath temperature on the entanglement. It shows that the entanglement is quite robust against all the dissipation rates of the system and bath temperature: it is present for the cavity, exciton, and mechanical decay rates being up to $\kappa_c/2\pi = {50}$ GHz, $\kappa_{1(2)}/2\pi = 1$ GHz, and $\kappa_b/2\pi = 10$ MHz under the parameters of Fig.~\ref{fig2}, which are within reach of current technology~\cite{review,EOM1,EOM3,Perrin13,Sesin}. It should, however, be noted that excessively high temperatures will break the binding of excitons. For instance, for GaAs materials with an exciton binding energy of $\sim$10 meV, the excitons will get ionized for temperatures higher than $\sim116$ K. Figure~\ref{fig4}(b) indicates that the excitonic entanglement would exist at $T>116$ K for materials with higher exciton binding energy.


At last, we show how to detect the generated exciton-exciton entanglement. The entanglement can be verified by measuring the CM of the two exciton modes. This can be realized by sending a {\it weak} probe field resonantly driving the cavity. Due to the exciton-photon state-swap interaction, the states of the two exciton modes are then mapped to the output field of the cavity. Therefore, by homodyning the cavity output field one can measure the quadratures of the exciton modes, based on which the CM is reconstructed. It should be noted that the output field carries the states of two exciton modes that differ by twice mechanical frequency in frequency, which can be extracted using two filters.  
This requires that the cavity photons decay much faster than the excitons do (i.e., $\kappa_c \gg \kappa_{1,2}$, which is adopted in our scheme), such that when the driving laser is switched off and all cavity photons die out, the exciton states remain almost unchanged, at which time the probe field is sent.

\section{Conclusions} 
\label{conc}

We present a theory for entangling two exciton modes in an exciton-optomechanics system, consisting of an optical cavity mode, a mechanical vibration mode, and two exciton modes. {The system is designed to have a strong exciton-phonon dispersive interaction while the optomechanical dispersive interaction is negligible.} 
The key idea is to simultaneously activate both the anti-Stokes and Stokes scattering, responsible for cooling the mechanical motion and creating the entanglement. {Due to the mediation of the mechanics in the above two scattering processes, the two exciton modes get entangled when they are respectively resonant with the two mechanical sidebands.}  The work enriches the quantum studies in exciton optomechanics and promotes the applications of the hybrid system in quantum information science and technology. 

{We note that this work significantly differs from~\cite{Zuo24}. Here, we entangle two independent exciton modes residing in two spatially separated quantum wells, and thus the entanglement embodies the nonlocal nature of the quantum correlation. By contrast, Ref.~\cite{Zuo24} entangles two components of the same exciton mode in a single quantum well in the strong-coupling regime forming exciton polaritons. This offers more degrees of freedom to enhance the tunability of the entanglement, e.g., the positions of the two quantum wells and the resonance frequencies of the two exciton modes can be independently adjusted.}  


\section*{Acknowledgments} 

This work has been supported by National Key Research and Development Program of China (Grant no. 2022YFA1405200) and National Natural Science Foundation of China (Grant no. 92265202).

\end{document}